\documentclass[aps,showpacs,preprintnumbers,nofootinbib]{revtex4}
\usepackage{graphicx}

\begin{document}
\preprint{gr-qc/yymmnnn}

\title{Is dark matter an extra-dimensional effect?}

\author{M. E. Kahil}
\email{kahil@aucegypt.edu} \affiliation{Mathematics Department,
Modern Sciences and Arts University, Cairo, Egypt}

\author{T. Harko}
\email{harko@hkucc.hku.hk} \affiliation{ Department of Physics and
Center for Theoretical and Computational Physics, The University
of Hong Kong, Pok Fu Lam Road, Hong Kong}

\date{\today}

\begin{abstract}
We investigate the possibility that the observed behavior of test
particles outside galaxies, which is usually explained by assuming
the presence of dark matter, is the result of the dynamical
evolution of particles in higher dimensional space-times. Hence,
dark matter may be a direct consequence of the presence of an
extra force, generated by the presence of extra-dimensions, which
modifies the dynamic law of motion, but does not change the
intrinsic properties of the particles, like, for example, the mass
(inertia). We discuss in some detail several possible particular
forms for the extra force, and the acceleration law of the
particles is derived. Therefore, the constancy of the galactic
rotation curves may be considered as an empirical evidence for the
existence of the extra dimensions.
\end{abstract}

\pacs{04.50.+h, 04.20.Jb, 04.20.Cv, 95.35.+d}

\maketitle




\section{Introduction}

There is a large amount of observational evidence showing that the standard
gravitational theories cannot describe correctly the large scale dynamics of
massive astrophysical systems. The observed rotational curves of spiral
galaxies cannot be explained by applying Newtonian or general relativistic
mechanics to the visible matter in galaxies and clusters. Neutral hydrogen
clouds are observed at large distances from the galactic center, much beyond
the extent of the luminous matter. A large number of independent
observations have shown that the rotational velocities $v_{tg}(r)$ of these
clouds tend toward constant and slightly rising values of the order of $%
v_{tg\infty }\sim 200-300$ km/s as a function of the distance $r$ from the
center of the galaxy \cite{Bi87,PeSaSt96,BoSa01}. This is in sharp contrast to the Newtonian
inverse square force law, which implies a decline in velocity. For clouds
moving in circular orbits with velocity $v_{tg}(r)$, the balance between the
centrifugal acceleration $v_{tg}^{2}/r$ and the gravitational attraction
force $GM(r)/r^{2}$ allows to express the mass-distance relation $M(r)$ in
the form $M(r)=rv_{tg}^{2}/G$. In the constant rotation velocity region this
leads to a mass profile $M(r)=rv_{tg\infty }^{2}/G$. Consequently, the mass
within a distance $r$ from the center of the galaxy increases linearly with $%
r$, even at large distances where very little luminous matter can be
detected.

This behavior of the galactic rotation curves is usually explained by
postulating the existence of some dark (invisible) matter, distributed in a
spherical halo around the galaxies. The dark matter is assumed to be a cold,
pressureless medium. There are many possible candidates for dark matter, the
most popular ones being the weakly interacting massive particles (WIMP) (for
a review of the particle physics aspects of dark matter see \cite{OvWe04}).
Their interaction cross section with normal baryonic matter, while extremely
small, are expected to be non-zero and we may expect to detect them
directly. It has also been suggested that the dark matter in the Universe
might be composed of superheavy particles, with mass $\geq 10^{10}$ GeV. But
observational results show the dark matter can be composed of superheavy
particles only if these interact weakly with normal matter, or if their mass
is above $10^{15}$ GeV \cite{AlBa03}.

Another interesting candidate for the dark matter is the dilatonic dark matter  - the fundamental scalar field which exists in all existing unified field theories.  The cosmological implications of the dilatonic dark matter have been explored in \cite{Cho90}, where a higher-dimensional generalization of the standard big-bang cosmology has been proposed. It was shown that the missing-mass problem as well as the horizon problem, and the flatness problem of the standard model can be resolved within the context of this unified cosmology.
The possibility that the dilaton plays the role of the dark matter of the universe was investigated in \cite{Cho3}. The condition for the dilaton to be the dark matter strongly restricts its mass to be around $0.5$ keV or $270$ MeV. For the other mass ranges, the dilaton contradicts the cosmological observations. The 0.5 keV dilaton has a free-streaming distance of about 1.4 Mpc and is an excellent candidate for warm dark matter, while the 270 MeV one has a free-streaming distance of about 7.4 pc and is a candidate for cold dark matter.  An experiment to detect the relic dilaton using the electromagnetic resonant cavity, based on the dilaton-photon conversion in strong electromagnetic background was proposed in \cite{Cho07}. The density of the relic dilaton, as well as an estimate of the dilaton mass for which the dilaton becomes the dark matter of the universe were calculated. The dilaton detection power in the resonant cavity were also obtained, and they were compared with the axion detection power in similar resonant cavity experiment. Based on the fact that the scalar curvature of the internal space determines the mass of the dilaton in higher-dimensional unified theories, the dilaton mass can explain the origin of the mass, and resolve the hierarchy problem \cite{Cho07a}. Moreover, cosmological observations put a strong constraint on the dilaton mass, and requires that the scale of the internal space to be larger than $10^{-9}$ m. Scalar fields or other long range
coherent fields coupled to gravity have also intensively been used to model
galactic dark matter \cite{scal1,scal2,scal3,scal4,scal5,scal6,scal7,Dah,Let1,Let2}.

However, despite more than 20 years of intense experimental and
observational effort, up to now no \textit{non-gravitational} evidence for
dark matter has ever been found: no direct evidence of it and no
annihilation radiation from it. Moreover, accelerator and reactor
experiments do not support the physics (beyond the standard model) on which
the dark matter hypothesis is based.

Therefore, it seems that the possibility that Einstein's (and the Newtonian)
gravity breaks down at the scale of galaxies cannot be excluded \textit{a
priori}. Several theoretical models, based on a modification of Newton's law
or of general relativity, have been proposed to explain the behavior of the
galactic rotation curves. A modified gravitational potential of the form $%
\phi =-GM\left[ 1+\alpha \exp \left( -r/r_{0}\right) \right] /\left(
1+\alpha \right) r,$with $\alpha =-0.9$ and $r_{0}\approx 30$ kpc can
explain flat rotational curves for most of the galaxies \cite{Sa84,Sa86}.

In an other model, called MOND, and proposed by Milgrom \cite{Mi1,Mi2,Mi3,Miin},
the Poisson equation for the gravitational potential $\nabla
^{2}\phi =4\pi G\rho $ is replaced by an equation of the form
$\nabla \left[ \mu \left( x\right) \left( \left| \nabla \phi
\right| /a_{0}\right) \right] =4\pi G\rho $, where $a_{0}$ is a
fixed constant and $\mu \left( x\right) $ a function satisfying
the conditions $\mu \left( x\right) =x$ for $x<<1$ and $\mu
\left( x\right) =1$ for $x>>1$. The force law, giving the acceleration $\vec{%
a}$ of a test particle, becomes $\mu \left( a/a_{0}\right)
\vec{a}=\vec{a}_{N}$, where $\vec{a}_{N}$ is the usual Newtonian acceleration. $a=a_{N}$ for $%
a_{N}>>a_{0}$ and $a=\sqrt{a_{N}a_{0}}$ for $a_{N}<<a_{0}$. The rotation
curves of the galaxies are predicted to be flat, and they can be calculated
once the distribution of the baryonic matter is known. The value of the
constant $a_{0}$ is given by $a_{0}\approx 1.2\times 10^{-8}\mathrm{cm}/%
\mathrm{s}^{-2}\approx cH_{0}/6\approx c\left( \Lambda /3\right) ^{1/2}/6$,
where $H_{0}$ is Hubble's constant and $\Lambda $ is the cosmological
constant. MOND is a purely phenomenological theory, but still it can explain
most of the galaxy rotation curves without introducing dark matter. But
despite its achievements, MOND has many problems of its own, like, for
example, the lack of conserved quantities, like energy, and a theoretical
justifications to the MOND phenomenology. A relativistic gravitation theory
for MOND was proposed by Bekenstein \cite{Be04}. In this model gravitation
is mediated by the tensor field $g_{\alpha \beta }$, a scalar field $\phi $
and a vector field $U_{\alpha }$, all three dynamical. For a simple choice
of its free function, the theory has a Newtonian limit for non-relativistic
dynamics with significant acceleration, but a MOND limit when accelerations
are small. A tensor-vector-scalar theory that reconciles the galaxy scale
success of modified Newtonian dynamics with the cosmological scale evidence
for cold dark matter (CDM) has been proposed by Sanders \cite{Sa05}. The
theory provides a cosmological basis for MOND by showing that the predicted
phenomenology only arises in a cosmological background.

Alternative theoretical models to explain the galactic rotation
curves have been elaborated recently by Mannheim \cite{Ma931,Ma932},
Moffat and Sokolov \cite{Mo96}, Brownstein and Moffat \cite{Br061,Br062}
and Roberts \cite{Ro04}. The constancy of the tangential velocity
of test particles orbiting around galaxies can be also explained
in the brane world models \cite{RS99a,RS99b}, where the effects of the projection of the  Weyl tensor from the bulk plays the role of the dark matter \cite{Ma041,Ma042,Ma043,Ma044,Ma045,Pal}, in the $f(R)$ modified
gravity models \cite{fR1,fR2,fR3,fR4}, and by assuming that dark matter is in
the form of a Bose-Einstein condensate \cite{Bo1,Bo2,Bo3,Bo4,Bo5,Bo6,Bo7}, or of an
Einstein cluster \cite{Ein1,Ein2}.

It is the purpose of the present paper to show that there is an
alternative general physical interpretation of the ''dark matter''
paradigm. More exactly, the constancy of the galactic rotation
curves can be obtained from the assumption that the motion of the
test particles in circular orbits around galaxies is non-geodesic.
The galactic dynamics of test particles is a direct consequence of
the presence of an extra force $f^{\mu } $, which modifies the
dynamic law of motion. Such a scenario, in which the galactic
rotation curves are explained by the presence of an extra force
may be called EFDOD (extra force dominated orbital dynamics).

One of the most interesting possibilities is that the extra force
is due to the presence of the extra dimensions. In such a model,
which we may call multidimensional EFDOD, the motion of the
particles takes place on geodesics in higher dimensions. A comprehensive geometric treatment of
Kaluza - Klein type unifications of non-Abelian gauge theories with gravitation was first
introduced in \cite{Cho1}, where the appearance of a cosmological constant was also noted,
and further developed in \cite{Cho1a,Cho1b}. The possible modifications of Einstein's theory of gravitation
due to the fifth force generated by the Kaluza - Klein dilaton were discussed in \cite{Cho2},
including the effects on the gravitational redshift, the deflection of light, the precession of perihelia, and the time-delay of radar echo around a spherically symmetric black hole in multidimensional space times. The long-range effect of the higher-dimensional fifth force is characterized by the dilatonic charge carried by the black hole even when it is neutral. In \cite{Cho92} it was emphasized that in the Brans-Dicke theory it is the Pauli metric, not the Jordan metric, which describes the massless spin-two graviton. Similarly, in the Jordan-Brans-Dicke theory, based on Kaluza-Klein unification, only the Pauli metric can correctly describe Einstein¡¦s theory of gravitation.
This necessitates a completely new reinterpretation of the  Kaluza-Klein cosmology, as well as of the
Brans-Dicke theory. More significantly, this analysis shows that the Kaluza-Klein dilaton must
generate a fifth force, which could violate the equivalence principle. Recent torsion-balance
experiments \cite{tors} have tested the gravitational inverse-square law at separations between 9.53 mm and 55 $\mu $m, thus probing distances less than the dark-energy length scale $d=85~\mu $m. It has been found with 95\% confidence that the inverse-square law holds down to a length scale  of around $56~\mu $m, and that an extra dimension must have a size $\leq 44~\mu $m.

However, it is known for some time that the effects of extra-dimensions on
the trajectory of test particles as observed in four dimensions
can be modeled in terms of an extra force $f^{\mu }$, for both
compactified and non-compactified spaces \cite{Cho3, Po01,Po04}. The presence of such a
force may explain the phenomenology and behavior of the galactic
rotation curves. We investigate in detail the dynamics of the test
particles in extra-dimensional models, and we find the conditions
which must be satisfied by the five-dimensional metric tensor in
order to explain the observed rotation curves. As a physical test
of our model we suggest that the lensing effects could be able to
find evidences for the multi-dimensional geometry.

We also investigate the acceleration law in EFDOD, and we find
that it has a striking similarity with the acceleration law in
MOND.  This leads us to the conclusion that the MOND theory, which
can be considered, from a physical point of view, as describing
the non-geodesic motion of a test particle in a gravitational
field under the action of an extra force generated by the
supplementary vector and scalar fields introduced in the model, is
a particular case of the EFDOD models.

The present paper is organized as follows. Physical models
determining a non-geodesic motion of particles are considered in
Section II. The behavior of test particles in stable circular
orbits in multidimensional models is considered in Section III. In
Section IV we consider the possibility that dark matter is an
extra-dimensional effect, and we obtain the general metric tensor
in the flat rotation curve region and we check the consistency of
the model. The acceleration law and the relation of our model with
MOND is discussed in Section V. We conclude and discuss our
results in Section VI. Throughout the paper we use the
Landau-Lifshitz conventions \cite{LaLi} for signature and metric.

\section{Scalar field generated extra force models}

There are several physical situations in which an extra force may
be present, determining a non-geodesic motion of the particles, like, for example, the
case of a real scalar field minimally coupled to gravity and
interacting with matter, the case of the extra-force generated by the non-trivial coupling between matter and geometry in the $f(R)$ modified gravity models, and the cases of the extra forces
generated by the presence of the compactified and non-compactified
higher dimensions of the space-time, respectively. In the present Section we will review briefly the first case.

In order to give a systematic treatment of the extra forces in the presence of a scalar field we will use the Bazanski approach for obtaining the geodesic equation \cite{Ba89}.
According to this approach, the equation of motion in any
dimensions can be obtained by applying the action principle to the
Lagrangian \cite{Ka06}
\begin{equation}
L=m(s)g_{AB}u^{A}\frac{D\Psi ^{B}}{Ds}+f_{A}\Psi
^{A},A,B=0,1,...,D,
\end{equation}
where $\Psi ^{B}$ is the deviation vector, $u^{A}$ is the tangent
vector to the geodesic and $f_{A}$ and $m(s)$ are functions which
depend on the specific physical models. The covariant derivative
$D\Psi ^{B}/Ds$ is defined as  $D\Psi ^{B}/Ds=d\Psi ^{B}/ds+\Gamma
_{CD}^{B}\Psi ^{C}u^{D}$.

The equation describing a real scalar field $\psi $ minimally
coupled to gravity and interacting with matter can be given as
\cite{Mb04}
\begin{equation}
\nabla _{\alpha }\nabla ^{\alpha }\psi =-J-\frac{\partial J}{\partial \psi }-%
\frac{\partial U}{\partial \psi },
\end{equation}
where $U=U\left( \psi \right) $ is the self-interaction potential and $%
J=J\left( x^{\beta }\right) $ is the source term of the scalar
field. In the following we neglect the effect of $U$ (assumed to
be of a breaking symmetry type). As for the source term $J$ we assume that it is of the general form $%
J=4\pi Gg(\psi )T_{\mu }^{\mu }/c^{2}$, where $T_{\mu }^{\mu }$ is
the trace of the energy momentum \ tensor of the matter and $g$ is
a coupling function satisfying the conditions $g\left( \psi
_{\infty }\right) =0$ and $\partial g\left( \psi _{\infty }\right)
/\partial \psi \neq 0$, respectively, where $\psi _{\infty }$ is
the value of the scalar field at the minimum of the potential. In
four dimensions the equation of motion of a test particle in the
presence of a scalar field can be derived from the Lagrangian
\begin{equation}
L=m(s)g_{\mu \nu }u^{\mu }\frac{D\Psi ^{\nu }}{Ds}+m(s)_{,\mu
}\Psi ^{\mu },
\end{equation}
and is given by
\begin{equation}
\frac{du^{\mu }}{ds}+\Gamma _{\alpha \beta }^{\mu }u^{\alpha }u^{\beta }=%
\frac{1}{m(s)}\left( g^{\mu \sigma }m_{,\sigma
}-\frac{dm}{ds}u^{\mu }\right) .
\end{equation}

By assuming that the
effective mass is of the form $m\sim \exp \left( -g\left( \psi \right) \psi \right) $ %
we obtain \cite{Mb04}
\begin{equation}
\frac{du^{\mu }}{ds}+\Gamma _{\alpha \beta }^{\mu }u^{\alpha }u^{\beta }=%
\frac{d\left( g\left( \psi \right) \psi \right) }{ds}u^{\mu
}-\partial ^{\mu }\left( g\left( \psi \right) \psi \right) .
\end{equation}

This equation of motion can also be derived from the variational principle $%
\delta \int mc\sqrt{g_{\mu \nu }u^{\mu }u^{\nu }}ds=0$. The force $\vec{f}$ has two components, one proportional to the velocity $%
\vec{v}$ of the particles, and which, being perpendicular to the
acceleration, does not give any contribution in Eq. (\ref{eq4})
and therefore can be neglected, and a second component given by $\vec{f}%
_{\parallel }=-\nabla \left[ g\left( \psi \right) \psi \right] $.
By assuming that the scalar field has a spherical symmetry, $\psi
=\psi \left( r\right) $, evaluating the force $\vec{f}_{\parallel
}$ around $\psi =\psi _{\infty }$ gives $f_{\parallel }\approx
-\left[ \partial g\left( \psi
_{\infty }\right) /\partial \psi \right] \psi ^{\prime }\psi $, where $%
^{\prime }$denotes the derivative with respect to $r$. In order to
obtain concordance with MOND, which means a constant $a_{0}$,  it
is necessary that $f_{\parallel }\sim -1/r$, which implies that
$\partial g\left( \psi _{\infty }\right) /\partial \psi =g^{\prime
}\left( \psi _{\infty }\right) >0$ and $\psi \left( r\right) =\psi _{0}%
\sqrt{\ln \left( r/R_{0}\right) }$, with $\psi _{0},R_{0}=$
constant.

As for the scalar field we assume that is satisfies the equation
\cite{Mb04}
\begin{equation}
\Delta \psi =\frac{4\pi G}{c^{2}} g^{\prime }\left( \psi _{\infty
}\right) \psi _{\infty }\rho ,
\end{equation}
where $\rho $ is the mass density of the matter fields other than
$\psi $. With the obtained form of the scalar field it follows
that the density of the matter interacting with the scalar field
has a density profile given by
\begin{equation}\label{dens}
\rho \left( r\right) =\frac{c^{2}}{16\pi G}\frac{\psi
_{0}}{g^{\prime
}\left( \psi _{\infty }\right) \psi _{\infty }} \frac{1}{r^{2}}%
\frac{8\ln ^{2}\left( r/R_{0}\right) +6\ln \left( r/R_{0}\right)
-1}{\ln ^{3/2}\left( r/R_{0}\right) }.
\end{equation}

Therefore EFDOD may be due to a scalar field interacting with a
matter distribution, whose density varies according to Eq.
(\ref{dens}). The presence of such a field could explain the
constancy of the galactic rotation curves and the corresponding
MOND phenomenology.

\section{Motion of test particles in stable circular orbits in
multidimensional space-times}

The above approach can naturally be implemented in Kaluza-Klein theory \cite{Cho90, Cho2}.  In the following we will consider the case of the extra forces generated by the
presence of the compactified and non-compactified higher dimensions of the
space-time.

Let the coordinates of the five-dimensional manifold, with metric tensor $%
\gamma _{AB}$, be $x^{A}$ $\left( A=0,1,2,3,4\right) $. The $5D$ interval is
given by $dS^{2}=\gamma _{AB}dx^{A}dx^{B}$. Usually it is assumed that the
first four coordinates $x^{\mu }$ are the coordinates of the space-time $%
x^{\mu }$ $\left( \mu =0,1,2,3\right) $, while $x^{4}=\xi $ is the
extra-dimension. Setting $\gamma _{\mu 4}=\gamma _{44}A_{\mu }$ and $\gamma
_{44}=\varepsilon \Phi ^{2}$, where $A_{\mu }$ and $\Phi $ are the vector
and scalar potentials, respectively, and $\varepsilon =\pm 1$, we may write
the line element without any loss of generality as \cite{Po01,Po04}
\begin{equation}
dS^{2}=g_{\mu \nu }dx^{\mu }dx^{\nu }+\varepsilon \Phi ^{2}\left( d\xi
+A_{\mu }dx^{\mu }\right) ^{2},
\end{equation}
where $g_{\mu \nu }=\gamma _{\mu \nu }-\varepsilon \Phi ^{2}A_{\mu }A_{\nu }$%
.

In the non-compact Kaluza-Klein theories, like, for example, the brane world
models \cite{RS99a}, all test particles travel on five dimensional
geodesics, but the observers, bounded to the usual four-dimensional
space-time, have access only to the $4D$ part of the trajectory.
Mathematically, this means that the equations governing the motion in $4D$
are projections of the $5D$ equations on the $4D$-hypersurfaces orthogonal
to some vector field $\psi ^{A}$. Generally, the background metric in $5D$
can be written as \cite{Po01,Po04, You00}
\begin{equation}
dS^{2}=\gamma _{\mu \nu }\left( x^{\alpha },\xi \right) dx^{\mu }dx^{\nu
}+\epsilon \Phi ^{2}\left( x^{\alpha },\xi \right) d\xi ^{2},
\end{equation}
where $\gamma _{\mu \nu }$ is the induced metric in $4D$. In brane world
theory the physical space time four dimensional metric $g_{\mu \nu }$ is
generally identified with $\gamma _{\mu \nu }$. However, in some approaches,
the physical metric in $4D$ is assumed to be conformally related to the
induced one,
\begin{equation}
dS^{2}=\Omega \left( \xi \right) g_{\mu \nu }\left( x^{\alpha },\xi \right)
dx^{\mu }dx^{\nu }+\epsilon \Phi ^{2}\left( x^{\alpha },\xi \right) d\xi
^{2}=\Omega \left( \xi \right) ds^{2}+\epsilon \Phi ^{2}\left( x^{\alpha
},\xi \right) d\xi ^{2},  \label{metr}
\end{equation}
where $\Omega \left( \xi \right) >0$ is called the warp factor \cite{Po04}.

In both compact and non-compact Kaluza-Klein theories, the motion of test
particles takes place in higher dimensions (usually the number of dimensions
of the space-time is assumed to be five), along the geodesics lines, and
with the equation of motion given by
\begin{equation}
\frac{du^{A}}{dS}+\hat{\Gamma}_{BC}^{A}u^{B}u^{C}=0.
\end{equation}
where $u^{A}=\left( dx^{\mu }/dS,d\xi /dS\right) $ is the five-velocity and $%
\hat{\Gamma}_{BC}^{A}$ are the Christoffel symbols formed with the $5D$
metric \cite{Cho2,Po01,Po04,Ka06}.

In order to obtain results which are relevant to the galactic dynamics, in
the following we will restrict our study to the static spherically-symmetric
five-dimensional metric given by
\begin{equation}
dS^{2}=e^{\nu \left( r,\xi \right) }c^{2}dt^{2}-e^{\lambda \left( r,\xi
\right) }dr^{2}-r^{2}\left( d\theta ^{2}+\sin ^{2}\theta d\phi ^{2}\right)
+\epsilon \Phi ^{2}\left( r,\xi \right) d\xi ^{2}=ds^{2}+\epsilon \Phi
^{2}\left( r,\xi \right) d\xi ^{2},  \label{metr1}
\end{equation}
where the coordinates have been chosen so that $x^{A}=\left( ct,r,\theta
,\phi ,\xi \right) $. The components of the five-velocity $U^{A}$ are given
by $U^{A}=dx^{A}/dS$. In particular, $U^{4}=d\xi /dS$. In the following we
also denote $u^{A}=dx^{A}/ds$, which represent the components of the
five-velocity with respect to the four-dimensional space-time with interval $%
ds$. The four-dimensional interval $ds$ is related to the five-dimensional
interval $dS$ by the relations $dS=ds/\sqrt{1-\epsilon \Phi ^{2}\left( r,\xi
\right) \left( U^{4}\right) ^{2}}$ or, equivalently, $dS=ds\sqrt{1+\epsilon
\Phi ^{2}\left( r,\xi \right) \left( u^{4}\right) ^{2}}$. The velocity $%
u^{A} $ is given as a function of $U^{A}$ by $u^{A}=U^{A}/\sqrt{1-\epsilon
\Phi ^{2}\left( r,\xi \right) \left( U^{4}\right) ^{2}}$.

The Lagrangian $L$ of a massive test particle traveling in the
five-dimensional space-time with metric given by Eq. (\ref{metr1}) is
\begin{equation}
2L=e^{\nu \left( r,\xi \right) }\left( \frac{cdt}{dS}\right)
^{2}-e^{\lambda \left( r,\xi \right) }\left( \frac{dr}{dS}\right)
^{2}-r^{2}\left[ \left( \frac{d\theta }{dS}\right) ^{2}+\sin ^2
\theta \left( \frac{d\phi }{dS}\right) ^{2}\right] +\epsilon \Phi
^{2}\left( r,\xi \right) \left( \frac{d\xi }{dS}\right) ^{2}.
\label{lag}
\end{equation}

Since the metric tensor coefficients do not explicitly depend on $ct$, $%
\theta $ and $\phi $, the Lagrangian (\ref{lag}) gives the following
conserved quantities (generalized momenta) in five dimensions:
\begin{equation}
e^{\nu \left( r,\xi \right) }\frac{cdt}{dS}=E={\rm const.},r^{2}\frac{%
d\theta }{dS}=L_{\theta }={\rm const.},r^{2}\sin ^{2}\theta \frac{d\phi }{dS%
}=L_{\phi }={\rm const.},  \label{cons}
\end{equation}
where $E$ is the total energy of the particle (in five-dimensions) and $%
L_{\theta }$ and $L_{\phi }$ are the components of the angular moment,
respectively. With the use of conserved quantities we obtain from Eq. (\ref
{metr1}) the geodesic equation for material particles as
\begin{equation}
e^{\nu +\lambda }\left( \frac{ds}{dS}\right) ^{2}\left( \frac{dr}{ds}\right)
^{2}+e^{\nu }\left[ 1+\frac{L_{T}^{2}}{r^{2}}-\epsilon \Phi ^{2}\left(
U^{4}\right) ^{2}\right] =E^{2},  \label{geod1}
\end{equation}
where we have denoted $L_{T}^{2}=L_{\theta }^{2}+L_{\phi }^{2}/\sin
^{2}\theta $. Eq. (\ref{geod1}) can be written as
\begin{equation}
e^{\nu +\lambda }\left( \frac{ds}{dS}\right) ^{2}\left( \frac{dr}{ds}\right)
^{2}+V_{eff}\left( r,\xi \right) =E^{2},
\end{equation}
where
\begin{equation}
V_{eff}\left( r,\xi \right) =e^{\nu }\left[ 1+\frac{L_{T}^{2}}{r^{2}}%
-\epsilon \Phi ^{2}\left( U^{4}\right) ^{2}\right] ,  \label{pot}
\end{equation}
is the effective potential of the motion, which also contains the effects of
the presence of the extra-dimension. If $\Phi \equiv 0$, we obtain the
well-known four-dimensional expression.

For the case of the motion of particles in circular and stable orbits the
generalized potential must satisfy the following conditions: a) $dr/ds=0$
(circular motion) b) $\partial V_{eff}/\partial r$ $=0$ (extreme motion) and
c) $\partial ^{2}V_{eff}/\partial r$ $^{2}!_{extr}>0$ (stable orbit),
respectively. Conditions a) and b) immediately give the conserved quantities
as
\begin{equation}
E^{2}=e^{\nu }\left[ 1+\frac{L_{T}^{2}}{r^{2}}-\epsilon \Phi ^{2}\left(
U^{4}\right) ^{2}\right] ,  \label{cons1}
\end{equation}
and
\begin{equation}
\frac{L_{T}^{2}}{r^{2}}=E^{2}\frac{r\nu ^{\prime }e^{-\nu }}{2}-\frac{r}{2}%
\frac{\partial }{\partial r}\left[ \epsilon \Phi ^{2}\left( U^{4}\right) ^{2}%
\right] ,  \label{cons2}
\end{equation}
respectively. Eqs. (\ref{cons1}) and (\ref{cons2}) allow us to express the
constants of the motion in the equivalent form
\begin{equation}
E^{2}=\frac{e^{\nu }}{1-\frac{r\nu ^{\prime }}{2}}\left\{ 1-\frac{1}{2r}%
\frac{\partial }{\partial r}\left[ r^{2}\epsilon \Phi ^{2}\left(
U^{4}\right) ^{2}\right] \right\} ,  \label{cons3}
\end{equation}
and
\begin{equation}
L_{T}^{2}=\frac{r^{3}\nu ^{\prime }}{2}\frac{1}{1-\frac{r\nu ^{\prime }}{2}}%
\left\{ 1-\frac{e^{-\nu }}{\nu ^{\prime }}\frac{\partial }{\partial r}\left[
e^{\nu }\epsilon \Phi ^{2}\left( U^{4}\right) ^{2}\right] \right\} ,
\label{cons4}
\end{equation}
respectively.

We define the tangential velocity $v_{tg}$ of a test particle in
four dimensions, measured in terms of the proper time, that is, by
an observer located at the given point, as \cite{LaLi}
\begin{equation}
v_{tg}^{2}=e^{-\nu }r^{2}c^{2}\left[ \left( \frac{d\theta }{cdt}\right)
^{2}+\sin ^{2}\theta \left( \frac{d\phi }{cdt}\right) ^{2}\right] =e^{-\nu
}r^{2}c^{2}\left[ \left( \frac{d\theta }{dS}\right) ^{2}+\sin ^{2}\theta
\left( \frac{d\phi }{dS}\right) ^{2}\right] \left( \frac{dS}{cdt}\right)
^{2}.
\end{equation}

By using the constants of motion from Eqs. (\ref{cons}) we immediately obtain
\begin{equation}
\frac{v_{tg}^{2}}{c^{2}}=\frac{L_{T}^{2}}{E^{2}}\frac{e^{\nu }}{r^{2}}.
\label{vtg}
\end{equation}

By eliminating the constant quantity $L_{T}^{2}/E^{2}$ between Eqs. (\ref
{cons3}) and (\ref{cons4}) gives
\begin{equation}
\frac{v_{tg}^{2}}{c^{2}}=\frac{\nu ^{\prime }r}{2}\frac{1-\frac{e^{-\nu }}{%
\nu ^{\prime }}\frac{\partial }{\partial r}\left[ e^{\nu }\epsilon \Phi
^{2}\left( U^{4}\right) ^{2}\right] }{1-\frac{1}{2r}\frac{\partial }{%
\partial r}\left[ r^{2}\epsilon \Phi ^{2}\left( U^{4}\right) ^{2}\right] }.
\label{vtg1}
\end{equation}

An alternative expression for the tangential velocity can be obtained
directly from the line element Eq. (\ref{metr1}), by using the constants of
motion. The result is
\begin{equation}
\frac{v_{tg}^{2}}{c^{2}}=\frac{e^{-\nu }E^{2}-1+\epsilon \Phi ^{2}\left(
U^{4}\right) ^{2}}{e^{-\nu }E^{2}}.
\end{equation}

With the use of Eq. (\ref{cons1}) we can express the tangential velocity as
\begin{equation}
\frac{v_{tg}^{2}}{c^{2}}=\frac{L_{T}^{2}}{L_{T}^{2}+r^{2}\left[ 1-\epsilon
\Phi ^{2}\left( U^{4}\right) ^{2}\right] }.  \label{vtg2}
\end{equation}

In order to completely solve the problem of the stable circular motion of
the test particles in extra-dimensional models we need an equation
determining $U^{4}$. This can be taken as the fifth component of the
geodesic equation, and is given by
\begin{equation}
\frac{d}{dS}\left( \epsilon \Phi ^{2}U^{4}\right) =\frac{1}{2}\frac{\partial
g_{\alpha \beta }}{\partial \xi }U^{\alpha }U^{\beta }.
\end{equation}

Taking into account that $U^{1}\equiv 0$, and that the $g_{22}$ and $g_{33}$
metric tensor components do not depend on $\xi $, we have
\begin{equation}
\frac{d}{dS}\left( \epsilon \Phi ^{2}U^{4}\right) =\frac{1}{2}\frac{\partial
g_{00}}{\partial \xi }U^{0}U^{0}=-\frac{1}{2}E^{2}\frac{\partial }{\partial
\xi }e^{-\nu },  \label{geod5}
\end{equation}
where we have again used the conservation law for the energy.

\section{Dark matter as an extra-dimensional effect}

The galactic rotation curves provide the most direct method of analyzing the
gravitational field inside a spiral galaxy. The rotation curves have been
determined for a great number of spiral galaxies. They are obtained by
measuring the frequency shifts $z$ of the light emitted from stars and from
the 21-cm radiation emission from the neutral gas clouds. Usually the
astronomers report the resulting $z$ in terms of a velocity field $v_{tg}$.
The observations show that at distances large enough from the galactic
center
\begin{equation}
v_{tg}\approx 200-300\text{ km/s}=\text{\textrm{constant.}}
\end{equation}

This behavior has been observed for a large number of galaxies \cite{Bi87}.

For a test particle traveling on a stable circular orbit in a
multi-dimensional space-time, the tangential velocity is given by Eq. (\ref
{vtg2}). In a purely four-dimensional space-time, $\Phi ^{2}\equiv 0$, and
the tangential velocity is given by $v_{tg}^{2}/c^{2}=L_{T}^{2}/\left(
L_{T}^{2}+r^{2}\right) $. In the limit of large $r$, $r\rightarrow \infty $,
and taking into account that $L_{T}^{2}$ is a finite (conserved) quantity, we obtain $%
\lim_{r\rightarrow \infty }v_{tg}=0$. However, the situation is quite
different in the multi-dimensional models. If the condition
\begin{equation}
1-\epsilon \Phi ^{2}\left( U^{4}\right) ^{2}=\frac{C^{2}\left( \xi \right)L_T^2}{%
r^{2}},  \label{cond1}
\end{equation}
holds true for large $r$, where $C^{2}\geq 0$ is an arbitrary
function of the fifth coordinate, then the tangential velocity of
a test particle in circular stable motion around the galactic
center is given by
\begin{equation}
\frac{v_{tg}^{2}}{c^{2}}=\frac{1 }{1 +C^{2}\left( \xi \right) }%
.
\end{equation}

If $C^{2}\left( \xi \right) $ is a true (galaxy-dependent)
constant, then the tangential velocity is an absolute constant,
too. Therefore, the constancy of the galactic rotation curves can
be explained naturally in the multi-dimensional physical models,
without the necessity of introducing the ad hoc hypothesis of the
dark matter. Of course, there are several other choices in Eq.
(\ref {vtg2}) which may lead to constant or slightly increasing
rotation velocity curves.

Since the tangential velocity is a constant or fifth-dimension
dependent quantity, one can solve Eq. (\ref{vtg1}) and find the
value of the metric tensor component $\exp \left( \nu \right) $ in
the constant rotational curves region. After some simple
transformations Eq. (\ref{vtg1}) can be written in the form
\begin{equation}
\nu ^{\prime }=2\frac{v_{tg}^{2}}{c^{2}}\frac{1}{r}+\left( 1-\frac{v_{tg}^{2}%
}{c^{2}}\right) \frac{1}{1-\epsilon \Phi ^{2}\left( U^{4}\right) ^{2}}\frac{%
\partial }{\partial r}\left[ \epsilon \Phi ^{2}\left( U^{4}\right) ^{2}%
\right] ,
\end{equation}
giving
\begin{equation}
e^{\nu }=D\left( \xi \right) r^{2v_{tg}^{2}/c^{2}}\left| \epsilon
\Phi ^{2}\left( U^{4}\right) ^{2}-1\right| ^{-\left(
1-v_{tg}^{2}/c^{2}\right) }, \label{enu}
\end{equation}
where $D\left( \xi \right) $ is an arbitrary integration function.
In the four-dimensional limit $\Phi ^{2}\equiv 0$ we obtain the well-known result $%
\exp \left( \nu \right) =Dr^{2v_{tg}^{2}/c^{2}}$, which has been
extensively used to discuss the properties of dark matter
\cite{scal1,scal2,scal3,scal4,scal5,scal6}, \cite{Ma041,Ma042,Ma043,Ma044}. Hence, the expression of the metric
tensor component $g_{00}$ in the constant rotation curves region
can be obtained in an exact form.

An important particular case corresponds to the situation in which $e^{\nu }$
is independent of $\xi $. Then, the geodesic equation Eq. (\ref{geod5}) can
be immediately integrated to give
\begin{equation}
\epsilon \Phi ^{2}U^{4}=B^2={\rm constant}.
\end{equation}

Together with Eq. (\ref{cond1}), the above first integral of the equations
of motion allows the determination of the functional form of the metric
tensor component $\Phi ^{2}$ in the constant rotation curves region as
\begin{equation}
\epsilon \Phi ^{2}=\frac{B^{2}}{1-C^2L_T^2/r^{2}}.
\end{equation}

In this case $B$, $C$ and $D$ are true constants, and they are
independent of the fifth dimension $\xi $. This model corresponds
to a compactified Kaluza-Klein type theory, in which all the
metric tensor components are independent of the fifth coordinate.

The multi-dimensional effects are dominant in the vacuum at large distances
from the galaxy. In the presence of matter, that it, inside or at the vacuum
boundary of the galaxy, these effects are very small, as compared to the
gravitational effect of the normal four-dimensional matter. Hence we may
assume that at the boundary of a galaxy with baryonic mass $M_{B}$ and
radius $R_{B}$ the four-dimensional geometry is approximately the
Schwarzschild geometry, and therefore
\begin{equation}
e^{\nu }|_{r=R_{B}}\approx 1-\frac{2GM_{B}}{c^{2}R_{B}}.
\end{equation}

This matching condition (approximately) determines the constant $D$ in Eq. (%
\ref{enu}).

\section{The acceleration law in EFDOD}

We start by assuming that the motion of a test particle in a $4D$ space-time
with metric $g_{\mu \nu }$ is given by
\begin{equation}
\frac{D^{(4)}u^{\mu }}{ds}\equiv \frac{du^{\mu }}{ds}+\Gamma _{\alpha \beta
}^{\mu }u^{\alpha }u^{\beta }=f^{\mu },  \label{eq1}
\end{equation}
where $u^{\mu }=dx^{\mu }/ds$ is the usual four-dimensional velocity of the
particle and $\Gamma _{\alpha \beta }^{\mu }$ are the Christoffel symbols
constructed by using the $4D$ metric. The presence of the extra force $%
f^{\mu }$ makes the motion of the particle non-geodesic. For $f^{\mu }\equiv
0$ we recover the geodesic equation of motion. All the usual gravitational
effects, due to the presence of an arbitrary mass distribution, are assumed
to be contained in the term $a_{N}^{\mu }=\Gamma _{\alpha \beta }^{\mu
}u^{\alpha }u^{\beta }$. In three dimensions and in the Newtonian limit, Eq.
(\ref{eq1}) can be formally represented as a three-vector equation of the
form
\begin{equation}
\vec{a}=\vec{a}_{N}+\vec{a}_f,  \label{eq2}
\end{equation}
where $\vec{a}$ is the total acceleration of the particle, $\vec{a}_{N}$ is
the gravitational acceleration and $\vec{a}_f$ is the acceleration (per unit
mass) due to the presence of the extra force. If $\vec{a}_f=0$, the equation
of motion is the usual Newtonian one, $\vec{a}=\vec{a}_{N}$, or,
equivalently, $\vec{a}=-GM\vec{r}/r^{3}$.

In the following we denote by $v^{2}=\vec{v}\cdot \vec{v}=\left| \vec{v}%
\right| ^{2}$ the magnitude of a vector $\vec{v}$. Taking the square of Eq. (%
\ref{eq2}) gives
\begin{equation}
\vec{a}_f\cdot \vec{a}_{N}=\frac{1}{2}\left( a^{2}-a_{N}^{2}-a_f^{2}\right) ,
\label{eq3}
\end{equation}
where $\cdot $ represents the three-dimensional scalar product. Eq. (\ref
{eq3}) can be interpreted as a general relation which gives the unknown
vector $\vec{a}_{N}$ as a function of the total acceleration $\vec{a}$, of
the acceleration $\vec{a}_f$ due to the extra force, and of the magnitudes $a^{2}$, $a_{N}^{2}$ and $%
a_f^{2}$, respectively. From Eq. (\ref{eq3}) one can express the vector $\vec{a%
}_{N}$, as one can easily check, in the form
\begin{equation}
\vec{a}_{N}=\frac{1}{2}\left( a^{2}-a_{N}^{2}-a_f^{2}\right) \frac{\vec{a}}{%
\vec{a}_f\cdot \vec{a}}+\vec{C}\times \vec{a}_f,  \label{eq4}
\end{equation}
where $\vec{C}$ is an arbitrary vector perpendicular to the vector $\vec{a}_f$%
. In the following, for simplicity, we assume $\vec{C}\equiv 0$.

The mathematical consistency of Eq. (\ref{eq4}) requires $\vec{a}_f\cdot \vec{a%
}\neq 0$, that is, the vectors $\vec{a}_f$ and $\vec{a}$ cannot be
perpendiculars. Generally, $\vec{a}_f\cdot \vec{a}=a_fa\cos \alpha $, where $%
\alpha $ is the angle between $\vec{a}_f$ and $\vec{a}$. Again, for
simplicity, we take $\alpha =0$, that is, we assume that the vectors $\vec{a}_f
$ and $\vec{a}$ are parallels. For $\vec{a}_f=0$, Eq. (\ref{eq4}) gives $%
a_{N}^{2}=a^{2}$, as required.

Therefore, we can represent the gravitational acceleration of a test
particle in the presence of an extra force as
\begin{equation}
\vec{a}_{N}=\frac{1}{2}\left( a^{2}-a_{N}^{2}-a_f^{2}\right) \frac{\vec{a}}{a_fa}%
.  \label{eq5}
\end{equation}

In the limit of very small gravitational accelerations $a_{N}<<a$, we obtain
the relation
\begin{equation}
\vec{a}_{N}\approx \frac{1}{2}a\left( 1-\frac{a_f^{2}}{a^{2}}\right) \frac{1}{a_f%
}\vec{a}.  \label{eq6}
\end{equation}

By denoting $(1/2a_f)\left( 1-a_f^{2}/a^{2}\right) =1/a_{E}$, Eq.
(\ref{eq6}) immediately gives
\begin{equation}
\vec{a}_{N}\approx \frac{a}{a_{E}}\vec{a},
\end{equation}
which is similar to the equation proposed phenomenologically by Milgrom \cite
{Mi1,Mi2,Mi3}. From this equation we obtain $a\approx \sqrt{a_{E}a_{N}}$, and since $%
a_{N}=GM/r^{2}$, we have $a\approx \sqrt{a_{E}GM}/r=v_{tg}^{2}/r$, where $%
v_{tg}$ is the rotational velocity of the test particle. Therefore, it
follows that $v_{tg}^{2}\rightarrow v_{\infty }^{2}=\sqrt{a_{E}GM}$, giving
the Tully-Fisher relation $v_{\infty }^{4}=a_{E}GM\sim L$, where $L$ is the
luminosity, assumed to be proportional to the mass \cite{Mi1,Mi2,Mi3}.

However, in the framework of EFDOD, $a_{E}$ is not generally a universal
constant, but it may be a position, acceleration or galaxy characteristics
dependent quantity.

Generally, from the given definition of $a_{E}$, we can
formally represent the extra acceleration as a function of $a$ and $a_{E}$ as $%
a_f/a_{E}=-a^{2}/a_{E}^{2}\pm \left( a/a_{E}\right) \sqrt{1+a^{2}/a_{E}^{2}}$.
Then, by means of some simple calculations, Eq. (\ref{eq5}) can be
represented as
\begin{equation}
\vec{a}_{N}=\frac{a}{a_{E}}\left[ F\left( \frac{a}{a_{E}}\right) \left(
\frac{a_{N}}{a}\right) ^{2}+1\right] \vec{a},
\end{equation}
where
\begin{equation}
F\left( \frac{a}{a_{E}}\right) =\frac{1}{2}\left( \frac{a}{a_{E}}\right)
^{-1}\left( \frac{a}{a_{E}}\mp \sqrt{1+\frac{a^{2}}{a_{E}^{2}}}\right) ^{-1}.
\end{equation}

With the use of the equation $a^{2}=v_{\infty
}^{4}/r^{2}=a_{E}GM/r^{2} $ we obtain the following expression for $a_{E}$:
\begin{equation}
a_{E}\approx \frac{a_f^{2}r^{2}}{GM}+2a_f.
\end{equation}

If $a_f\sim GMa_0 /r$, where $a_0$ is a constant, then in the large $r$
limit, when $a_f\rightarrow 0$, $a_{E}\approx a_0 ^{2}$ is a constant,
whose numerical value is determined by the physical properties of the extra
force. If the extra-force is universal in its nature, than $a_{E}$ is a
universal constant.

Therefore the MOND paradigm (which also postulates the existence of a
universal constant $a_{0}$) is equivalent, and can be derived, from the
assumption of the non-geodesic motion of the test particles around the
galactic centers under the action of a specific force.

\section{Discussions and final remarks}

In the present paper we have shown that the dynamics of the test particles
in circular orbits around galaxies can be attributed to the presence of an
extra force, generated by the presence of the extra dimensions, which modifies the standard (Newtonian or general relativistic)
motion, by giving a supplementary contribution to the acceleration. The
total acceleration has a general form, which is formally identical to the one
proposed on a phenomenological-empirical basis in MOND. In an equivalent
formulation, MOND is the result of the non-geodesic motion of particles
under the influence of a specific force. On the other hand, depending on the
physical nature of the extra force, more general physical models than MOND
can be obtained. We have considered several possibilities for the extra
force. The extra force may be generated, for example, by a scalar field
coupled with matter. However, in such a scenario, some extra (dark?) matter
is required, and, in order to obtain a constant $a_0$, a specific density
profile for the matter is necessary. On the other hand,  for the scalar field to influence the galactic rotational velocity, it has to be massless. But generally massless scalar fields could not exist in nature.

However, one could relate extra dimensions to dark matter through the dilatonic dark matter \cite{Cho90,Cho92}. As an essential part of higher dimensional metric the dilaton plays a crucial role
to determine the higher dimensional geodesic. But this higher dimensional
geodesic equation, expressed in the lower dimension, contains
a non-geodesic force, created by the dilaton, which requires a modification of general relativity.
Generally, in $4D$ the motion can be described as taking place under the
effect of a tensor (metric) field and of a vector and scalar field,
respectively. Interestingly enough, the relativistic version of MOND \cite
{Be04} requires exactly such a modification of general relativity, but with
the extra-fields introduced by hand. However, for this model to be true, it is necessary that the extra- force from the extra dimension has to be long ranged (in a galactic scale). On the other hand the fifth force obtained in the framework of the compactified Kaluza-Klein theories cannot be long ranged \cite{Cho3}. Some possibilities of overcoming these difficulties may be by assuming that the extra-dimensions are large, as is the case in the brane-world models \cite{RS99a,Po01,Po04}.

In order to explore in more detail the connections between EFDOD, MOND and
dark matter, some explicit physical models are necessary to be built. This
will be done in some forthcoming papers.

\acknowledgments

We would like to thank to the anonymous referee for comments and suggestions that helped us to significantly improve the manuscript. The work of T. H. was supported by the GRF grant No. 7018/08P of the government of the Hong Kong SAR.

\end{document}